\documentclass[twocolumn,superscriptaddress,floatfix,preprintnumbers, nofootinbib,hyperref]{revtex4} 
\pdfoutput=1
\usepackage[colorlinks=true,breaklinks=true]{hyperref}
\usepackage[utf8]{inputenc}
\hypersetup{allcolors=[rgb]{0.0 0.0 0.6},linkcolor=[rgb]{0.75 0.05 0.05}}
\usepackage{amsmath,amssymb,amsfonts}
\usepackage{epsfig}  
\usepackage{float}
\usepackage{graphicx}   
\usepackage{slashed}  
\usepackage{tabulary}
\usepackage{url}
\usepackage{color}
\usepackage{multirow}
\usepackage[normalem]{ulem}
\usepackage[dvipsnames]{xcolor}
\usepackage{letltxmacro}
\LetLtxMacro{\oldcite}{\cite}
\renewcommand{\cite}[1]{\mbox{\oldcite{#1}}}

\newcommand{\beq}{\begin{equation}}
\newcommand{\eeq}{\end{equation}}

\allowdisplaybreaks

\bibpunct{[}{]}{,}{n}{}{,}

\begin{document}

\title{Robust marginalization of baryonic effects for cosmological inference at the field level}

\author{Francisco Villaescusa-Navarro}\email{villaescusa.francisco@gmail.com}
\affiliation{Department of Astrophysical Sciences, Princeton University, Peyton Hall, Princeton NJ 08544, USA}
\affiliation{Center for Computational Astrophysics, Flatiron Institute, 162 5th Avenue, New York, NY, 10010, USA}

\author{Shy Genel}
\affiliation{Center for Computational Astrophysics, Flatiron Institute, 162 5th Avenue, New York, NY, 10010, USA}
\affiliation{Columbia Astrophysics Laboratory, Columbia University, New York, NY, 10027, USA}

\author{Daniel Angl\'es-Alc\'azar}
\affiliation{Department of Physics, University of Connecticut, 196 Auditorium Road, Storrs, CT, 06269, USA}
\affiliation{Center for Computational Astrophysics, Flatiron Institute, 162 5th Avenue, New York, NY, 10010, USA}

\author{David N. Spergel}
\affiliation{Center for Computational Astrophysics, Flatiron Institute, 162 5th Avenue, New York, NY, 10010, USA}
\affiliation{Department of Astrophysical Sciences, Princeton University, Peyton Hall, Princeton NJ 08544, USA}

\author{Yin Li}
\affiliation{Center for Computational Astrophysics, Flatiron Institute, 162 5th Avenue, New York, NY, 10010, USA}

\author{Benjamin Wandelt}
\affiliation{Sorbonne Universite, CNRS, UMR 7095, Institut d’Astrophysique de Paris, 98 bis boulevard Arago, 75014 Paris, France}
\affiliation{Center for Computational Astrophysics, Flatiron Institute, 162 5th Avenue, New York, NY, 10010, USA}

\author{Leander Thiele}
\affiliation{Department of Physics, Princeton University, Princeton, NJ 08544, USA}

\author{Andrina Nicola}
\affiliation{Department of Astrophysical Sciences, Princeton University, Peyton Hall, Princeton NJ 08544, USA}

\author{Jose Manuel Zorrilla Matilla}
\affiliation{Department of Astrophysical Sciences, Princeton University, Peyton Hall, Princeton NJ 08544, USA}

\author{Helen Shao}
\affiliation{Department of Astrophysical Sciences, Princeton University, Peyton Hall, Princeton NJ 08544, USA}

\author{Sultan Hassan}
\affiliation{Center for Computational Astrophysics, Flatiron Institute, 162 5th Avenue, New York, NY, 10010, USA}
\affiliation{Department of Physics \& Astronomy, University of the Western Cape, Cape Town 7535, South Africa}

\author{Desika Narayanan}
\affiliation{Department of Astronomy, University of Florida, Gainesville, FL, USA}
\affiliation{University of Florida Informatics Institute, 432 Newell Drive, CISE Bldg E251, Gainesville, FL, USA}

\author{Romeel Dave}
\affiliation{Institute for Astronomy, University of Edinburgh, Royal Observatory, Edinburgh EH9 3HJ, UK}
\affiliation{Department of Physics \& Astronomy, University of the Western Cape, Cape Town 7535, South Africa}
\affiliation{South African Astronomical Observatories, Observatory, Cape Town 7925, South Africa}

\author{Mark Vogelsberger}
\affiliation{Kavli Institute for Astrophysics and Space Research, Department of Physics, MIT, Cambridge, MA 02139, USA}

\date{\today}
\smallskip


\begin{abstract}
We train neural networks to perform likelihood-free inference from $(25\,h^{-1}{\rm Mpc})^2$ 2D maps containing the total mass surface density from thousands of hydrodynamic simulations of the CAMELS project. We show that the networks can extract information beyond one-point functions and power spectra from all resolved scales ($\gtrsim 100\,h^{-1}{\rm kpc}$) while performing a robust marginalization over baryonic physics at the field level: the model can infer the value of $\Omega_{\rm m} (\pm 4\%)$ and $\sigma_8 (\pm 2.5\%)$ from simulations completely different to the ones used to train it.
\end{abstract}

\maketitle


The wealth and quality of data from upcoming surveys such as DESI \cite{DESI_2016}, eRosita \cite{Predhel_2021}, Euclid \cite{Laureijs_2011}, PFS \cite{Takada_2014}, SKA \cite{Taylor_1999}, Roman \cite{Spergel_2015}, and Rubin \cite{LSST_2009}
is expected to bring cosmology to a different level. The data from these surveys contain the key to answering fundamental questions in physics such as what the nature of dark energy or the sum of the neutrino masses are. The main obstacle to achieving this is not the amount and quality of data, but two significant theoretical challenges: 1) the optimal summary statistic(s) needed to extract the maximum amount of information from data in the mildly and fully non-linear regimes is unknown, and  2) astrophysical processes such as feedback from supernovae and active galactic nuclei (AGN) are expected to impact the non-linear regime in a poorly understood manner.

Ideally, one would want to use an estimator that is able to extract the maximum amount of information directly from the field while marginalizing over baryonic effects at the same time. In \cite{Paco_2020b} we demonstrated with toy examples that neural networks can achieve both goals. In our companion paper \citep{baryons_marginalization} we have shown that neural networks can extract cosmological information embedded in state-of-the-art hydrodynamic simulations from complex 2D fields such as gas temperature, electron number density, and magnetic fields, while marginalizing over baryonic effects at the same time.

\begin{figure*}
\centering
  \includegraphics[width=0.99\linewidth]{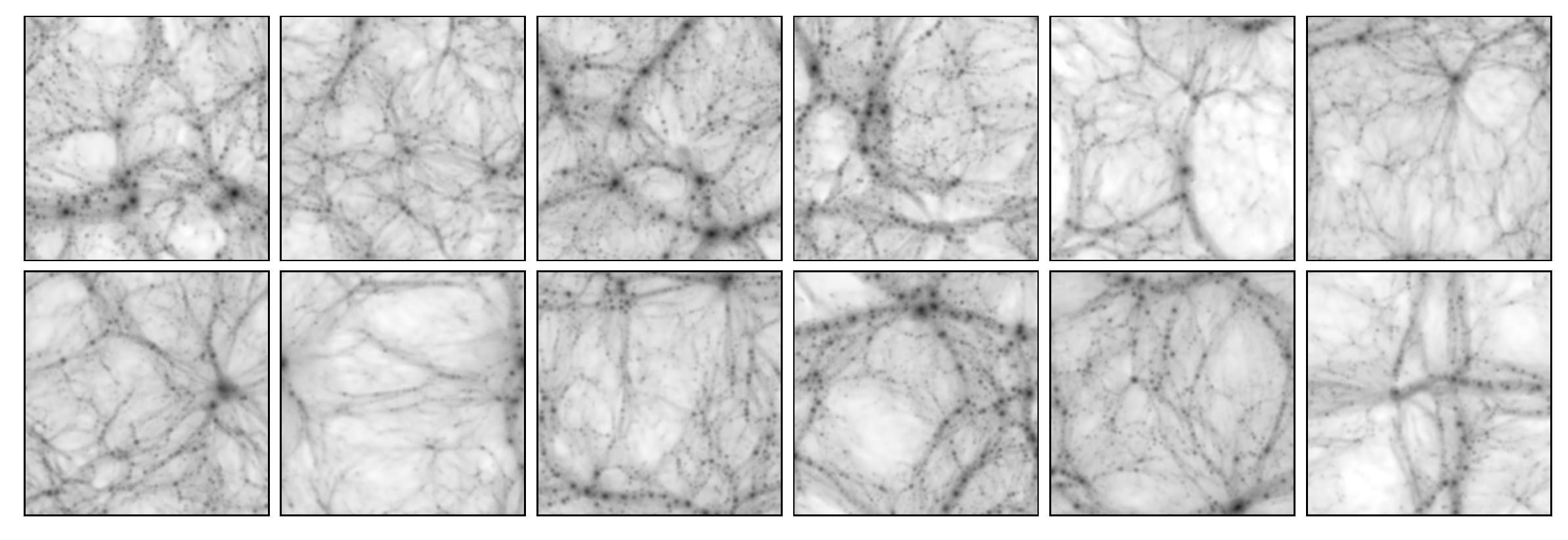}
  \vspace{-0.3cm}
\caption{Examples of the projected total mass maps used to train the neural networks from the IllustrisTNG (top row) and SIMBA (bottom row) simulations. Each map contains $256\times256$ pixels and has a physical size of $25\times25~(h^{-1}{\rm Mpc})^2$.}
\label{fig:maps}
\end{figure*}

In this letter we show, for the first time, that this procedure can be robust. In other words, it does not rely on training or testing the model on a particular type of simulation. We will show that neural networks trained on total matter surface density maps from one suite of simulations can infer the correct cosmology even when tested on maps from another suite, and vice versa. It is not obvious that this should work, because the two suites follow a different set of hydrodynamic equations, use distinct numerical methods to solve them, and employ different subgrid models. We also show that rather than simply extracting cosmological information only from large scales to avoid bias, our networks successfully extract cosmological information from the smallest scales resolved in the maps. These results have important consequences for weak lensing surveys, paving the way to a method that can extract all information from the field while marginalizing over baryonic effects.

We use data from the Cosmology and Astrophysics with MachinE Learning Simulations (CAMELS) project \citep{CAMELS} at $z=0$. CAMELS contains two simulation suites, IllustrisTNG and SIMBA, which have been run using the AREPO \cite{Arepo} and GIZMO \cite{Hopkins2015_Gizmo} codes, employing the subgrid physics implemented in the IllustrisTNG \citep{WeinbergerR_16a,PillepichA_16a} and SIMBA \citep{SIMBA} simulations, respectively. For each suite, we use the latin-hypercube (LH) set, which contains 1,000 simulations, each with a different value of the cosmological parameters $\Omega_{\rm m}$ and $\sigma_8$, the fractional matter density and the amplitude of fluctuations in spheres of radius 8 $h^{-1}{\rm Mpc}$, and four astrophysical parameters controlling the efficiency of supernova and AGN feedback ($A_{\rm SN1}$, $A_{\rm SN2}$, $A_{\rm AGN1}$, $A_{\rm AGN2}$) as well as a different value of the random seed used to generate the initial conditions. The value of the cosmological and astrophysical parameters is varied across a very wide range; e.g. $\Omega_{\rm m}\in[0.1, 0.5]$ and $\sigma_8\in[0.6, 1.0]$. We refer the reader to \cite{CAMELS} for further details on the CAMELS simulations.

From these simulations we generate 2D total mass surface density maps as follows. First, we take slices of dimensions $25\times25\times5~(h^{-1}{\rm Mpc})^3$ and project all particle positions in the slice into the 2D plane along the third axis. We then assign the particle positions and their masses to a 2D regular grid with $256\times256$ pixels, considering that each particle represents a circle with uniform density within its radius. For gas and dark matter particles this radius is set to the distance from the particle to its 32nd nearest neighbor, while stars and black holes are considered to be point masses. Finally, we divide the projected mass in every pixel by its area to obtain the total mass surface density. For each simulation we produce 15 maps, giving rise to 15,000 maps for the IllustrisTNG simulations, and another 15,000 maps for the SIMBA simulations. In Fig. \ref{fig:maps} we show a few examples of these maps. We provide further details on the method used to generate the maps in our companion paper \citep{CMD}.

We use these maps to train moment neural networks \cite{Moment_networks} to infer the values of the marginal posterior mean ($\mu_i$) and standard deviation ($\sigma_i$) for each parameter $\theta_i$. The output of the network is thus twelve numbers, two per parameter. The loss function we optimize via gradient descent is
\begin{eqnarray}
\mathcal{L}&=&\sum_{i=1}^6\log\left(\sum_{j\in{\rm batch}}\left(\theta_{i,j} - \mu_{i,j}\right)^2\right)\nonumber \\
+&&\sum_{i=1}^6\log\left(\sum_{j\in{\rm batch}}\left(\left(\theta_{i,j} - \mu_{i,j}\right)^2 - \sigma_{i,j}^2 \right)^2\right)~,
\end{eqnarray}
where the interior sum runs over all maps in the batch while the external sum runs over all six parameters. 

Our architecture consists in a set of 6 blocks, where each block follows the structure CBACBACBA, where C, B, and A are convolutional, batchnorm, and LeakyReLU layers, respectively.
In each block, the first two convolutional layers have kernel size of 3, stride 1 and padding 1, while the last layer has kernel size of 2, padding 0 and stride 2. The first convolutional layer of the first block is not followed by a batchnorm layer. After the six blocks, we use a smaller block with CBA where the convolutional layer has kernel size 4, stride 1 and padding 0. The output of the last block is flattened and passed into two fully connected layers. Our models have between 10 and 30 million free parameters. 

We train the networks for 200 epochs using the AdamW optimizer \citep{AdamW} with batch size equal to 128. We perform hyper-parameter optimization using the optuna package \citep{Optuna}. We train using data augmentation on the maps, rotations and flipping of the maps, to teach the model to preserve rotational and parity symmetries. We emphasize that by construction, our model is already invariant under translations. We provide further details on the architecture and training procedure in our companion paper \citep{CMD}. 

\begin{figure*}
\centering
  \includegraphics[width=0.99\linewidth]{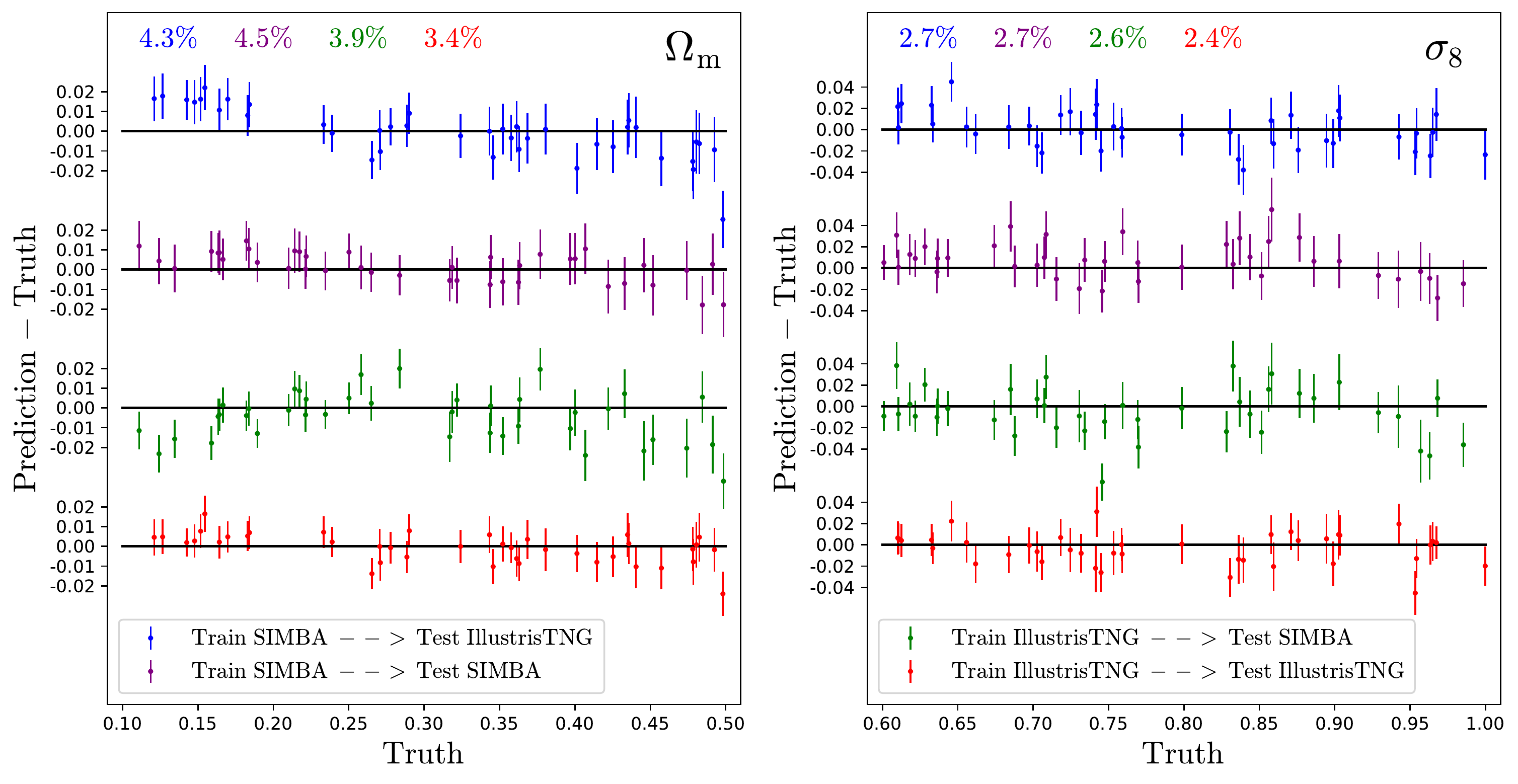}
  \vspace{-0.3cm}
\caption{We train two neural networks to perform likelihood-free inference on the value of the cosmological and astrophysical parameters. The two networks are trained using either IllustrisTNG or SIMBA total mass maps from CAMELS simulations at $z=0$. We then test the models using either IllustrisTNG or SIMBA maps from their test sets. We show the results for $\Omega_{\rm m}$ and $\sigma_8$ in the left and right panels, respectively. The dots with errorbars represent the posterior mean and standard deviation from a single map. We have subtracted the true value to the posterior mean to facilitate
visualization.
The network is capable to recover the true value of $\Omega_{\rm m}$ and $\sigma_8$ with high-accuracy in all situations, even when the network is tested on maps from simulations completely different to the ones used to train the model.}
\label{fig:TNG2SIMBA}
\end{figure*} 

We first train and validate a network using maps from 900 and 50 IllustrisTNG simulations, respectively. We test the model on maps from the 50 IllustrisTNG simulations of the test set (Fig. \ref{fig:TNG2SIMBA}). We find that the model is able to infer the value of $\Omega_{\rm m}$ and $\sigma_8$ with high accuracy: $\langle\delta \Omega_{\rm m}/\Omega_{\rm m}\rangle\simeq3.4\%$, $\langle\delta \sigma_8/\sigma_8\rangle\simeq2.4\%$. The network can also infer the value of $A_{\rm SN1}$ and $A_{\rm SN2}$ but with much larger errors: $\langle\delta A_{\rm SN1}/A_{\rm SN1}\rangle\simeq38\%$ and $\langle\delta A_{\rm SN1}/A_{\rm SN1}\rangle\simeq17\%$. On the other hand, the network is not capable of constraining the parameters controlling the efficiency of AGN feedback. 

We now test the above model using maps from the SIMBA simulations. We emphasize that the above model has never seen a map from those simulations in the training. We show the results in Fig. \ref{fig:TNG2SIMBA} for a subset of 40 maps. Each point represents a map from the SIMBA simulations; the dot is the posterior mean while the errorbar shows the posterior standard deviation. We find that the network trained on the IllustrisTNG maps is able to infer the true value of $\Omega_{\rm m}$ and $\sigma_8$ from the SIMBA maps with a similar accuracy to that for the IllustrisTNG maps: $\langle\delta \Omega_{\rm m}/\Omega_{\rm m}\rangle\simeq3.9\%$, $\langle\delta \sigma_8/\sigma_8\rangle\simeq2.6\%$. We emphasize that the SIMBA simulations not only solve the hydrodynamic equations in a different way than the IllustrisTNG simulations, but the subgrid model is also quite distinct. Thus, it is not obvious that this test should work at all. On the other hand, the network completely fails at predicting the astrophysical parameters. This is expected since the astrophysical parameters in the IllustrisTNG and SIMBA simulations are different in definition and in their effect on multiple quantities (see \citep{CAMELS} for a detailed discussion). 

Next, we train another network with the same architecture but using maps from the SIMBA simulations. As above, we employ maps from 900 and 50 simulations for training and validating the model. We first test the model using SIMBA maps from the test set. We find results very similar to those obtained when training and testing with IllustrisTNG maps: the network is able to infer the value of $\Omega_{\rm m}$ and $\sigma_8$ with high accuracy, $\langle\delta \Omega_{\rm m}/\Omega_{\rm m}\rangle\simeq4.5\%$, $\langle\delta \sigma_8/\sigma_8\rangle\simeq2.7\%$ (see Fig. \ref{fig:TNG2SIMBA}). The model can also put some constraints on $A_{\rm SN1}$ ($49\%$) and $A_{\rm SN2}$ ($23\%$), while the AGN parameters cannot be inferred. We note that the average errors on the value of the cosmological parameters from this network are slightly higher than those from the model trained on IllustrisTNG maps. This could be because the SIMBA simulations have a wider range of effective variation in the astrophysics parameters, which could force the network to marginalize more deeply than in the IllustrisTNG maps.

Finally, we test this model using IllustrisTNG maps. The results are very similar to those discussed above. Namely, we find that the model is able to accurately infer the value of the cosmological parameters: $\langle\delta \Omega_{\rm m}/\Omega_{\rm m}\rangle\simeq4.3\%$, $\langle\delta \sigma_8/\sigma_8\rangle\simeq2.7\%$. On the other hand, this model is unable to infer any value of the astrophysical parameters of these maps, as expected for the reasons discussed above.

These results illustrate how neural networks can 1) extract information from the field and 2) marginalize over baryonic effects. The most important point we would like to emphasize in this paper is that this procedure seems to be robust, i.e., it does not rely on training on a particular set of simulations with a particular hydrodynamics solver and implementation of subgrid physics. In our companion paper \citep{baryons_marginalization} we showed that this may not always be the case; when repeating the above exercise using gas temperature maps, the model is not robust.

We now carry out a set of tests as an attempt to understand what the network is doing and what it is not. First, we investigate whether the network is only using information from the total mass in the maps to do the inference. Since the total mass in a simulation is proportional to $\Omega_{\rm m}$, we expect a certain degree of correlation between that parameter and the total mass of our maps. We first make a scatter plot between the total mass in a map and its value of $\Omega_{\rm m}$ and $\sigma_8$, and as expected we find a correlation with $\Omega_{\rm m}$. However, it is not large enough to explain the few percent errors we obtain from the network. This occurs because our maps only represent volumes of $25\times25\times5~(h^{-1}{\rm Mpc})^3$ and so cosmic variance breaks the perfect correlation between $\Omega_{\rm m}$ and the total mass in a simulation box. Furthermore, with the network trained, we feed it with maps where all pixels contain the mean mass density per pixel. In this case, the network is not able to infer the value of any of the parameters. This shows that the network is not just using the total mass in a map to do the inference. 

Next, we explore whether the network is performing inference via clustering or just through some 1-point function. To this end, we input maps where the pixel values are randomly shuffled into the network; if the network was just extracting information from, e.g., the probability distribution function, then the ordering of the pixels would not matter. We find that also in this case the network cannot perform parameter inference. This shows that the network is using information from the spatial distribution of the maps.

We also study whether the network is extracting more information than the traditional summary statistic used in cosmology: the power spectrum. For this purpose, we compute the power spectrum of all the maps and perform likelihood-free inference on it. We use fully connected layers and perform hyper-parameter optimization over the number of layers, number of neurons per layer, learning rate, weight decay, and dropout rate. We find that the information embedded in the power spectrum only suffices to infer the value of $\Omega_{\rm m}$ with very large error bars: $\langle\delta {\Omega}_{\rm m}/\Omega_{\rm m}\rangle\simeq20\%$. For the other five parameters the network only predicts the mean with large errorbars. This test shows that the network is extracting information beyond the power spectrum; that information beats the one in the power spectrum by a large margin. 

One hypothesis that could explain our results is that the contamination from baryonic effects in the maps is negligible, so the network does not need to learn to marginalize over baryonic effects. This could also explain why training on one simulation type and testing on another works so well. To test this hypothesis, we train a neural network on maps from the CAMELS N-body simulations, which span the same range in $\Omega_{\rm m}$ and $\sigma_8$ as their hydrodynamic counterparts. These simulations do not model baryonic effects, so the network does not need to marginalize over them. We input maps of the total mass from the hydrodynamic simulations into the network trained on the N-body maps. We find that the network fails to infer the value of the cosmological parameters. This clearly shows that our maps are sensitive to baryonic effects. This is completely expected, and we refer the reader to Fig. 7 of \citep{CAMELS} for a plot showing the effect of baryons on the 3D matter power spectrum at $z=0$.

One may wonder if the neural network is just marginalizing over baryonic effects by ignoring scales affected by them; one can achieve this by, e.g., convolving the maps with a filter that suppresses clustering below a given scale. If that was the case, then the network trained on total mass maps from hydrodynamic simulations would also work when testing on maps from N-body simulations. When we make this test we also find that the network cannot perform a good inference on the value of the cosmological parameters, showing that the network does not seem to be marginalizing over baryonic effects by ignoring small scales.

To demonstrate more clearly that the network is extracting cosmological information from all scales, we first smooth the total mass maps from the hydrodynamic simulations using a Gaussian kernel with different widths of one, two, and three pixels. We then retrain the model with these three types of maps. We find that these models are able to infer the value of the cosmological parameters, but the errorbars are larger; the larger the width of the Gaussian kernel, the larger the standard deviation of the posterior. 
This clearly indicates that the original network, trained on maps that were not smoothed,  was extracting cosmological information from the smallest scales, $\sim 100~h^{-1}{\rm kpc}$, the ones that are expected to be more heavily contaminated by baryonic effects. We provide further details on these tests in the supplemental material.

From these tests we can draw the following conclusions: 1) the network is not doing something trivial such as using the total mass in the maps or computing its 1-point PDF, but 2) the network is extracting information beyond the power spectrum on different scales, including the smallest scales sampled by our maps, while 3) the network is marginalizing over baryonic effects at the field level. More importantly, even if the methods used to solve the hydrodynamic equations in the IllustrisTNG and SIMBA simulations are very different, as are their subgrid models, one can train networks using one set of simulations that are able to marginalize over baryonic effects when testing on the other set. This shows how robust and powerful this method can potentially be.

We leave for future work a deeper and more complete analysis of what features and elements of the cosmic web the network is focusing its attention on in order to perform the inference. We will also investigate in more detail how the network marginalizes over baryonic effects and extracts cosmological information from the smallest scales sampled by our maps. It is important to understand why this method seems robust against astrophysics and how this information can be leveraged with real surveys. 

We will also study how robust parameter inference is with other fields, like the ones we have explored in \cite{baryons_marginalization} such as gas mass, gas temperature, and electron density. We emphasize that the results reported in this work are probably lower bounds, as real data may be affected by systematics and noise, which will increase errors. On the other hand, our method is able to extract robust information from very small scales from fields that are highly non-linear, showing that it should be easier to apply this procedure to real data, where projections will Gaussianize weak lensing maps. On the other side, from the smoothing analysis above, we expect there would be significantly less information for the network to extract.

Our results have important implications for cosmology. For example, while the 2D total mass density projections we have considered in this paper are not directly observable, they can be seen as the lensing planes that ultimately give rise to weak gravitational lensing.  Our framework opens the door to extracting information from such maps down to the smallest resolved scales, which are rich in information.  Furthermore, we have shown that one can simulate baryonic effects using a particular method to solve the hydrodynamic equations and subgrid implementation, and as long as the range of variation is large enough, be robust to the exact scheme employed. Overall, this work illustrates the potential of applying deep learning to optimally tackle parameter estimation in cosmology from upcoming multi-wavelength surveys.
  
\textit{Acknowledgments---} 
The training of the neural networks has been carried out using GPUs from the \textit{Rusty} and \textit{Tiger} clusters at the Flatiron Institute and Princeton University, respectively. FVN acknowledges funding from the WFIRST program through NNG26PJ30C and NNN12AA01C. DAA was supported in part by NSF grants AST-2009687 and AST-2108944. The work of SG, DAA, YL, SH, BW, and DNS has been supported by the Simons Foundation. The maps used for this paper are part of the CAMELS Multifield Dataset (CMD), that we present and make publicly available in our companion paper \citep{CMD}. Instructions on how to download the data, together with the scripts and neural networks weights can be found in \url{https://camels-multifield-dataset.readthedocs.io}. We have made use of the \textit{Pylians} libraries, publicly available at \url{https://pylians3.readthedocs.io}. Details about the CAMELS simulations can be found in \url{https://www.camel-simulations.org}. 

\vspace{0.5cm}


\twocolumngrid
\begin{center}
{ \it \large Supplemental Material}\\
\end{center}

Here we provide further details on some of the tests carried out that due to space limitations were only briefly described in the main text.

\begin{figure*}
\centering
  \includegraphics[width=0.36\linewidth]{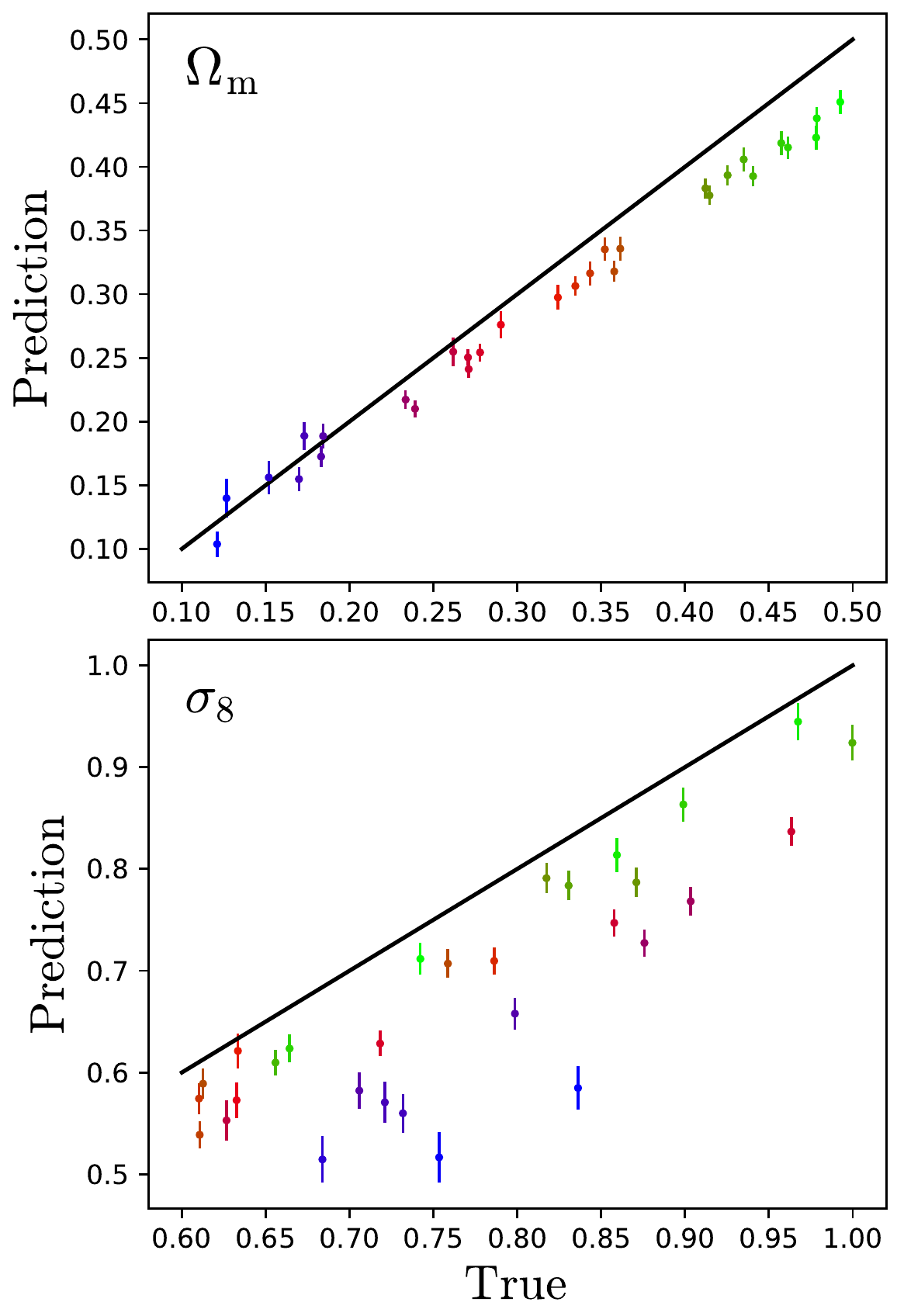}
  \hspace{2cm}
  \includegraphics[width=0.36\linewidth]{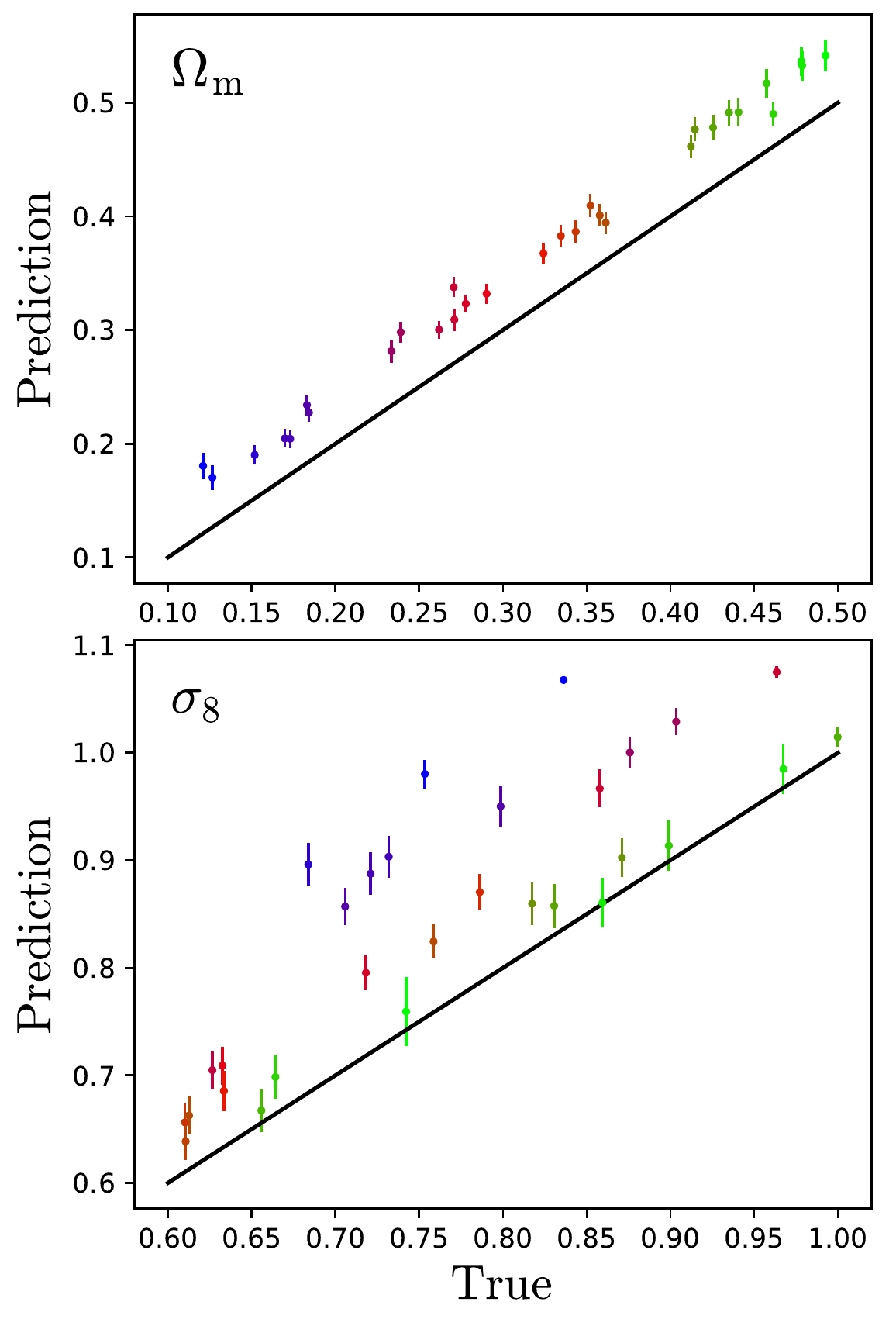}
  \vspace{-0.3cm}
\caption{\textbf{Left:} We have trained a neural network using maps from N-body simulations and tested the model on maps from the total matter mass density of IllustrisTNG simulations. \textbf{Right:} We have tested the model that was trained on total matter mass density maps from IllustrisTNG simulations on N-body maps. As can be seen, neither test worked, indicating that 1) the reason why the networks perform so well cannot be that maps are barely affected by baryonic effects and 2) the network is not applying a cut on the scales affected by baryonic effects.}
\label{fig:Nbody}
\end{figure*}

\textbf{Baryonic effects}. As discussed in the main text, one of the potential explanations of our results is that the maps we are considering have a negligible contamination by baryonic effects. Thus, it does not really matter whether they are produced from IllustrisTNG or SIMBA simulations, and a network trained on one should work on the other. If that would be the case, we could train a neural network on maps from the N-body simulations, that are not contaminated by baryonic effects and then test it on maps from either IllustrisTNG or SIMBA. The left panel of Fig. \ref{fig:Nbody} shows the results of training a network on maps from the N-body simulations and testing them on total matter mass density from the IllustrisTNG simulations. As can be seen, the network is not able to infer the value of neither $\Omega_{\rm m}$ nor $\sigma_8$. It is however interesting to observe that in the case of $\Omega_{\rm m}$, the network is almost working; the prediction exhibits a bias that gets larger with higher values of $\Omega_{\rm m}$. We do not attempt to explain these results here.

\textbf{Cutoff scale.} In cosmological analyses it is a standard practice to apply a cutoff in scale in order to avoid uncertainties in non-linear scales as well as baryonic effects. We may wonder if our network is doing something similar, i.e. extracting information from scales not affected by baryonic effects while neglecting all information from scales affected by astrophysics. If that would be the case, we would expect that a network trained on maps from the hydrodynamic simulations would be able to infer the value of $\Omega_{\rm m}$ and $\sigma_8$ from maps from N-body simulations. We have carried out this test and show the results in the right panel of Fig. \ref{fig:Nbody}. As can be seen, also in this case the network is not able to infer the value of the cosmological parameters from the N-body maps. It is also very interesting to note that the prediction of $\Omega_{\rm m}$ seems to be off by a constant factor. We believe this offset has some physical implications, although we do not attempt to provide an explanation in this work. For $\sigma_8$ instead, we find that the prediction is worse for lower values of $\Omega_{\rm m}$; we found a similar behavior in the test mentioned above.

\begin{figure*}
\centering
  \includegraphics[width=0.99\linewidth]{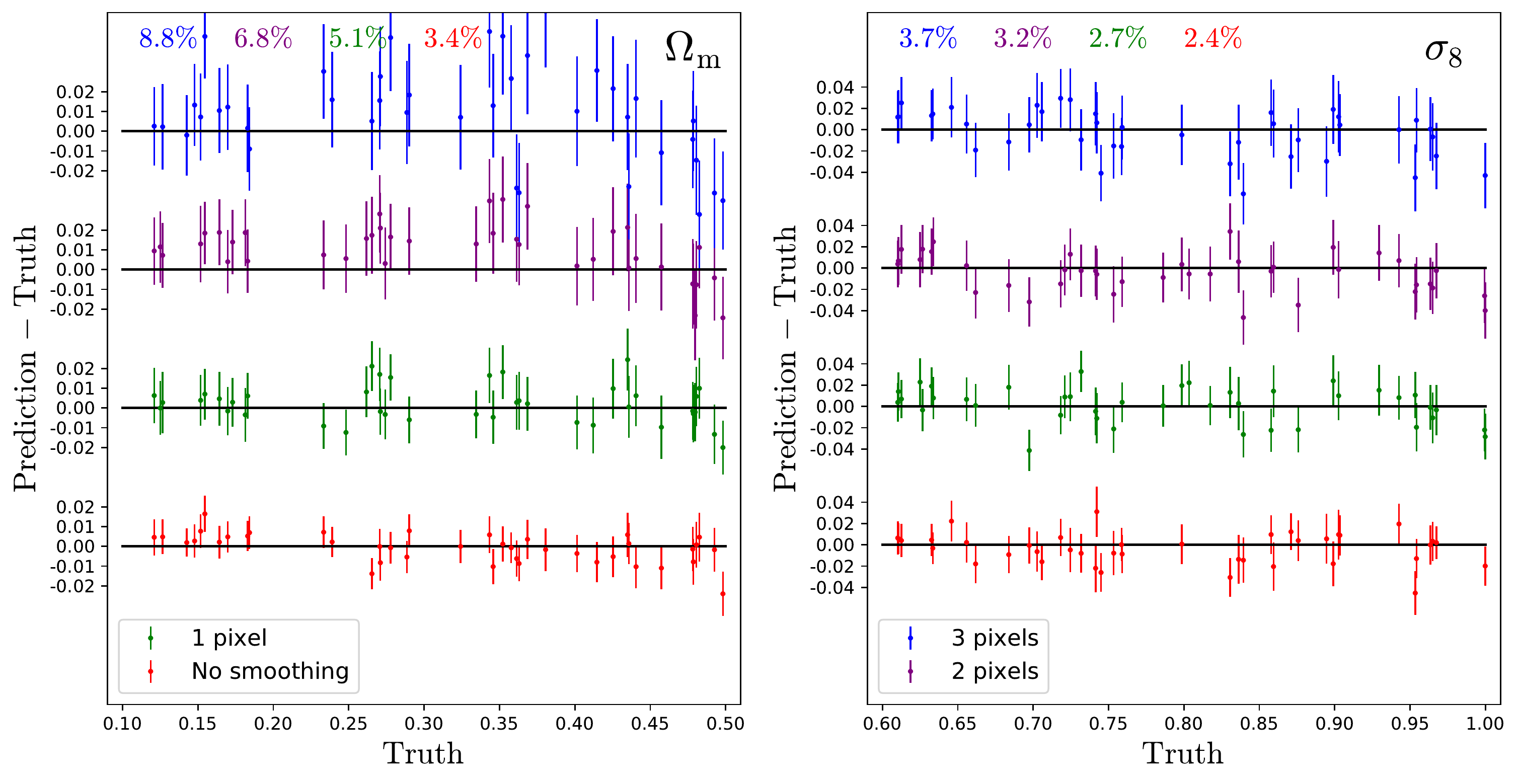}
  \vspace{-0.3cm}
\caption{We have trained four neural networks on IllustrisTNG maps with no smoothing (red) and smoothing the maps with a Gaussian kernel of size 1 pixel (green), 2 pixels (magenta), and 3 pixels (blue). This figure shows the results when testing these networks on IllustrisTNG maps with the same characteristic as those used in their training. The average relative accuracy, $\langle \sigma_i/\mu_i\rangle$, is displayed in the upper part of the plot. As can be seen, by smoothing the maps, the network is only able to infer the value of the cosmological parameters with lower accuracy. This indicates that the network trained on the maps that are not smoothed is extracting information from the smallest scales (1 pixel) available in the map.}
\label{fig:smoothing}
\end{figure*}

\textbf{Smoothing test.} In order to investigate on which scales the network is extracting information, we have taken the IllustrisTNG maps and smooth them with a Gaussian kernel of width equal to 1 pixel, 2 pixels, and 3 pixels. Those maps are then used to train three different neural networks. In Fig. \ref{fig:smoothing} we show the results. As can be seen, the larger the width of the Gaussian kernel, the less information the network is able to extract. The average relative accuracy on $\Omega_{\rm m}$ degrades from 3.4\% in the case of no smooth to 5.1\% (1 pixel), 6.8\% (2 pixels), and 8.8\% (3 pixels); for $\sigma_8$ it changes from 2.4\% in the case of no smoothing to 2.7\% (1 pixel), 3.2\% (2 pixels), and 3.7\% (3 pixels). These results indicate that the original network is extracting cosmological information from the smallest scales resolved in our maps, i.e. at the pixel level. Besides, this also suggest that more information can be extracted from higher resolution maps.

\bibliography{Refs}
\bibliographystyle{utcaps}
\bigskip

\end{document}